\documentclass{article}
\usepackage{spconf,amsmath,graphicx}
\usepackage{physics}
\usepackage{esvect}
\usepackage{floatrow}
\usepackage{acronym}
\usepackage[acronym]{glossaries}
\newcommand{\etal}{\emph{et al.} }
\newcommand{\eg}{\emph{e.g.,} }

\newcommand{\fref}[1]{Figure \ref{#1}}		%Referencing figures
\newcommand{\eref}[1]{Eq. \eqref{#1}}

\newacronym{strf}{STRF}{spectro-temporal receptive field}
\newacronym{stmi}{STMI}{spectro-temporal modulation index}
\newacronym{stmr}{STMR}{spectro-temporal modulation response}
\newacronym{stme}{STME}{spectro-temporal modulation error}
\newacronym{se}{SE}{speech enhancement}
\newacronym{dnn}{DNN}{deep neural network}
\newacronym{tf}{TF}{time-frequency}
\newacronym{tfe}{TFE}{time-frequency error}
\newacronym{mse}{MSE}{mean-squared error}
\newacronym{stftm}{STFTM}{short-time Fourier transform magnitude}
\newacronym{sid}{SID}{speaker identification}
\newacronym{lps}{LPS}{log power spectra}
\newacronym{stft}{STFT}{short-time Fourier transform}
\newpage

\title{A Modulation-Domain Loss for Neural-Network-based Real-time Speech Enhancement}
\name{Tyler Vuong,$^{1}$
      Yangyang Xia,$^{1}$
      Richard M. Stern$^{1,2}$}
      
\address{$^1$ Department of Electrical and Computer Engineering, Carnegie Mellon University\\
         $^2$ Language Technologies Institute, Carnegie Mellon University \\
       \texttt{tvuong@andrew.cmu.edu}, \texttt{raymondxia@cmu.edu}, \texttt{rms@cs.cmu.edu}\\   
}

\begin{document}

   © 2021 IEEE.  Personal use of this material is permitted.  Permission from IEEE must be obtained for all other uses, in any current or future media, including reprinting/republishing this material for advertising or promotional purposes, creating new collective works, for resale or redistribution to servers or lists, or reuse of any copyrighted component of this work in other works

%\ninept
%
\maketitle
\begin{abstract}
We describe a modulation-domain loss function for deep-learning-based speech enhancement systems. Learnable spectro-temporal receptive fields (STRFs) were adapted to optimize for a speaker identification task. The learned STRFs were then used to calculate a weighted \acrfull{mse} in the modulation domain for training a speech enhancement system. Experiments showed that adding the modulation-domain MSE to the MSE in the spectro-temporal domain substantially improved the objective prediction of speech quality and intelligibility for real-time speech enhancement systems without incurring additional computation during inference.
\end{abstract}
\begin{keywords}
Real-time speech enhancement, spectro-temporal receptive field, loss functions
\end{keywords}
\section{Introduction}
\label{sec:intro}
Supervised \gls{se} using deep neural networks (DNNs) has received tremendous attention in recent years. The availability of abundant amounts of training data and  advancements in \acrshort{dnn} architectures have resulted in systems that provide better performance than  the ideal binary mask \cite{luoConvTasNetSurpassingIdeal2019} -- a target that highly correlates with speech intelligibility \cite{loizouReasonsWhyCurrent2011,kryterValidationArticulationIndex1962}. The core design of a   DNN-based SE system involves  decisions that perform compensation in one feature domain and calculate the loss function in another. Despite the emerging time-domain compensation methods (\eg \cite{luoConvTasNetSurpassingIdeal2019}), predicting a time-varying gain function, or a \gls{tf} mask  \cite{wangTrainingTargetsSupervised2014}, has been the most popular and reliable approach.

Loss functions for supervised \gls{se} in the \gls{tf} domain have historically been   calculated in the time or frequency domain. However, most existing loss functions \cite{braunConsolidatedViewLoss2020} were motivated by statistically-optimal solutions \cite{ephraimSpeechEnhancementUsing1984,ephraimSpeechEnhancementUsing1985,limEnhancementBandwidthCompression1979} and do not necessarily correlate with perceptual quality or intelligibility of speech \cite{loizouSpeechEnhancementTheory2013}. More recently, perceptually-motivated loss functions have been proposed to optimize modified predictors of speech quality \cite{martin-donasDeepLearningLoss2018} and intelligibility \cite{zhaoPerceptuallyGuidedSpeech2018}. Interestingly, these methods did not show improvement over objective metrics for which the loss functions did not directly optimize, suggesting that there is room for improving their generalization ability.

Modulation is closely related to  speech intelligibility. The speech transmission index (STI)  measures the extent to which amplitude modulation of speech is preserved in degraded environments and is highly correlated with speech intelligibility \cite{steeneken1980physical,paytonMethodDetermineSpeech1999}.  The \gls{stmi} was subsequently proposed to account for joint spectro-temporal modulation \cite{elhilaliSpectrotemporalModulationIndex2003}. \gls{se} in the modulation domain has also been explored \cite{mesgaraniDenoisingDomainSpectrotemporal2007,mirbagheriNonlinearFilteringSpectrotemporal2010}. However, these methods assume that speech and noise are separable in the modulation domain. Moreover, they typically require a complete set of spectro-temporal receptive fields (STRFs)  in order to invert the processed modulation spectra back to the \gls{tf} domain.  This may  be computationally infeasible for real-time applications.

In this paper, we propose a simple \gls{mse}-based loss function in the spectro-temporal modulation domain for supervised \gls{se}. We call the loss \gls{stme} because of its close relation to template-based \gls{stmi} \cite{elhilaliSpectrotemporalModulationIndex2003}, which correlates well with speech intelligibility. The calculation of the \gls{stme} is based on a set of pre-selected modulation kernels, which had been shown to be critical for the accuracy of predicted speech intelligibility using speech stimuli \cite{paytonMethodDetermineSpeech1999}. Following our recent success in discriminating live speech from synthetic or broadcast speech using learnable \gls{strf} kernels \cite{vuong2020learnable}, we develop an automatic way to determine these kernels through an auxiliary \gls{sid} task. \gls{stme} is applicable to any \gls{dnn} system as long as the \gls{stftm} of the target and degraded speech are accessible when training the \gls{dnn}. Our proposed system's loss is easy to compute, does not incur additional computation during inference, and avoids lossy inversion, which is a problem with conventional modulation-domain SE approaches    \cite{mesgaraniDenoisingDomainSpectrotemporal2007}.

\textbf{Organization of this paper.}
In the next section, we introduce related work in supervised \gls{se} and the background of \gls{stmi}. We then describe our selection procedure for the \gls{strf}s and the loss function. Finally, we describe the evaluation procedure and discuss the experimental results.

%RELATED WORK STUFF
%Related work on strfs and learnable wavelet-STRFs -> we use learnable Gabor-based STRFs
%Related work on estimating ratio mask in spectral-temoral modulation domain either by using a VAD to construct ideal mask(no training data for learning) or by solving optimization an problem that estimates the ratio mask that minimizes the distance between clean STMD and enhanced STMD (optimization problem is solved for each new file and still uses a VAD for speech and noise frames) -> we estimate ratio mask in the TF domain using a DNN ( more efficient, smaller architecture than a network that estimates the mask in STMD - size will scale with # of kernels if we estimate a mask in STMD) and then we optimize the distance between clean STMD and enhanced STMD during TRAINING (similar to STMI).  at testing, its efficient and this method makes no changes to the architecure and adds no additional latency during testing for real-time applications. 

\section{BACKGROUND}

\label{sec:prior}
In this section we briefly review supervised speech enhancement and the spectro-temporal modulation index.

\textbf{Supervised \gls{dnn}-based speech enhancement.}
We assume that the observed noisy speech contains clean speech corrupted by additive noise. This relationship in the \gls{stft} domain is described by 
\begin{equation}\label{eqn:STFT}
X[t,k] = S[t,k] + N[t,k]
\end{equation}
where $X[t,k], S[t,k],$ and $N[t,k]$ represent the STFT at time frame $t$ and frequency bin $k$ of the observed noisy speech, clean speech and noise, respectively. Without loss of generality, we assume that a \gls{dnn} is trained to predict a magnitude gain, $G[t, k]$, from past and current information of degraded speech. The enhanced \gls{stftm} is obtained by element-wise multiplication of  the predicted gain by the noisy \gls{stftm},
\begin{equation} \label{eqn:IRM prediction}
\left|\widehat{S}[t,k]\right|= G[t,k]\left|X[t,k]\right|.
\end{equation}
A popular loss function that drives the learning for this \gls{dnn} is the \gls{mse} between the enhanced STFTM and the clean STFTM, or the \gls{tfe},
\begin{equation}\label{eqn:mse_loss}
\text{TFE} = ||(\Vec{S} - \Vec{\widehat{S}})||^{2}
\end{equation}
where $\Vec{S}$ and $\Vec{\widehat{S}}$ denote the vector representations of the clean STFTM and enhanced STFTM, respectively.

\textbf{Spectro-temporal modulation index.}
The spectro-temporal modulation index (STMI) is a measure of speech integrity in the modulation domain as viewed by a model of the auditory system \cite{elhilaliSpectrotemporalModulationIndex2003}. At the core of this model is a bank of \gls{strf}s that are believed to exist in the central auditory system and respond to a range of patterns of temporal and spectral modulation \cite{chiMultiresolutionSpectrotemporalAnalysis2005}.  Each \gls{strf} is a \gls{tf} template in which two seed functions control the temporal modulation (rate) and spectral modulation (scale) selectivity, respectively. The \gls{stmr} of a spectrographic representation of speech to a \gls{strf} is defined by the convolution of the two,
\begin{equation} \label{eqn:stmr}
    \text{STMR}_S[t,k] = \text{STRF}[t,k] * f(|S[t,k]|^2)
\end{equation}
 where $f$ is a frequency-integration function followed by a logarithmic compression that mimics the function of the early auditory system. Finally, the template-based \gls{stmi} \cite{elhilaliSpectrotemporalModulationIndex2003} is defined as
\begin{equation}
    \text{STMI}^{\text{T}} = 1 - \frac{||\overrightarrow{\text{STMR}}_S - \overrightarrow{\text{STMR}}_X||^2}{||\overrightarrow{\text{STMR}}_{S}||^2},
\end{equation}
where $\text{STMR}[t,k]$ is typically integrated over time before being converted to the vector form. In previous implementations of  modulation-domain \gls{se}, compensation was performed directly on $\text{STMR}_X[t,k]$ \cite{mesgaraniDenoisingDomainSpectrotemporal2007,mirbagheriNonlinearFilteringSpectrotemporal2010}. In our method, we perform enhancement in the \gls{tf} domain and use a loss function in the modulation domain that is closely related to $\text{STMI}^{\text{T}}$ for training the \gls{dnn}. 

It should be noted that the selection of meaningful modulation frequency becomes an issue when speech signals (instead of modulated noise) are used to calculate an estimate of speech intelligibility \cite{paytonMethodDetermineSpeech1999}. Previous work using \gls{strf}s typically performed dimensionality reduction on features extracted from densely-sampled \gls{strf}s \cite{mesgaraniDiscriminationSpeechNonspeech2006,meyerRobustnessSpectrotemporalFeatures2011,ravuriEasyDoesIt2012}. We learn the parameters of the \gls{strf}s through an auxiliary \gls{sid} task. We describe our own method next.

\section{Method}
In this section, we present the DNN system we used for speech enhancement and the calculation of our spectral-temporal modulation error loss.

\textbf{Speech enhancement system.}
We used the normalized \gls{lps} as the input feature.  The \gls{stft} is first obtained using a 20-millisecond Hamming window with 50\% overlap and a 512-point discrete Fourier transform.  Then we take the natural logarithm of the power of the \gls{stft} and normalize the \gls{lps} with frequency-dependent online normalization following \cite{xiaWeightedSpeechDistortion2020}.

To estimate the magnitude gain for each frame, we used a similar real-time network architecture to the one described in \cite{braunConsolidatedViewLoss2020}.  The network consists of a single fully connected (FC) layer followed by two stacked unidirectional Gated Recurrent Units (GRUs) and three more FC layers.  Rectified linear unit (ReLU) activation is used after each of the FC layers except the very last one where a sigmoid activation is used to bound the output magnitude gain to be between zero and one.  We obtain the enhanced waveform by multiplying the magnitude gain element-wise with the noisy STFTM and using the original noisy phase for reconstruction.  In total, the network contained roughly 2.8 million learnable parameters. 

\begin{figure}[htb]
\centering
\includegraphics[width = \textwidth]{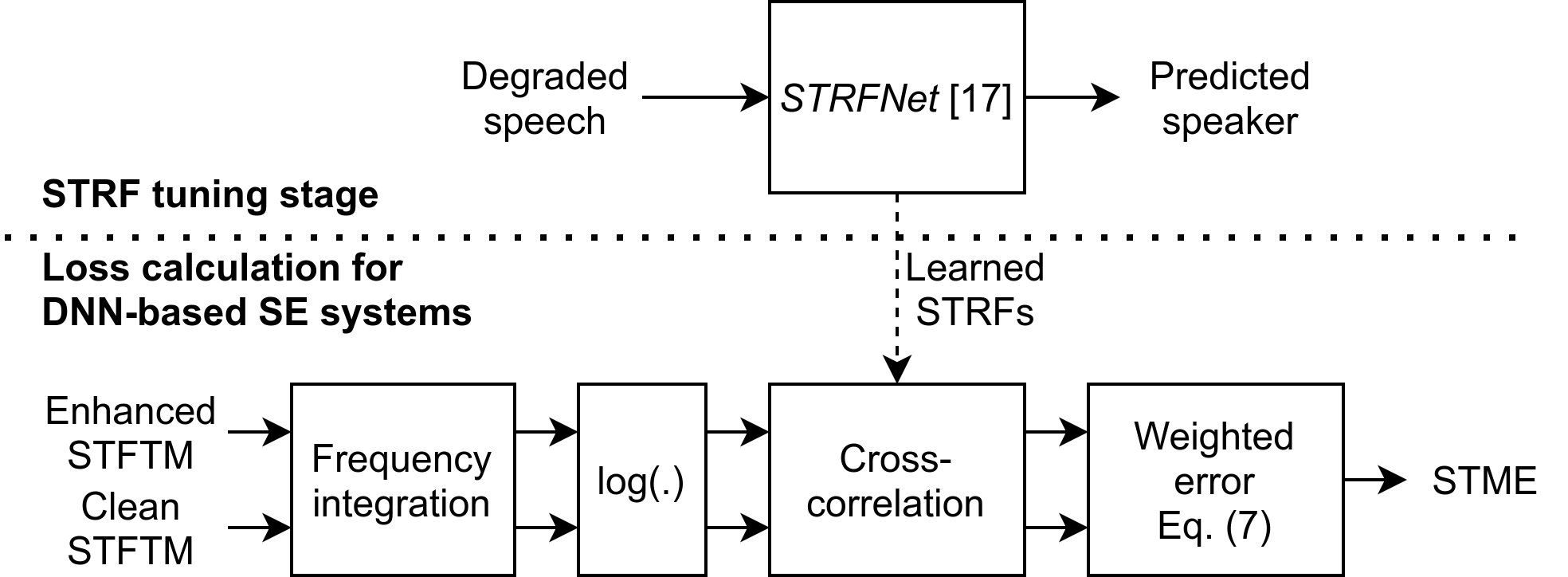}
\caption{Flow diagrams of the \gls{strf} tuning stage (top) and the \gls{stme} loss calculation stage (bottom).}
\label{fig:stme}
\vspace{-4mm}

\end{figure}

\textbf{Tuning learnable \gls{strf}s on speaker identification.}
One central problem involving the construction of the loss function is the selection of modulation parameters that are relevant to speech intelligibility \cite{paytonMethodDetermineSpeech1999}. Previous work has shown that the \gls{stmr} is redundant \cite{mesgaraniDiscriminationSpeechNonspeech2006} and the possible values for those modulation parameters span a wide range \cite{meyerRobustnessSpectrotemporalFeatures2011,ravuriEasyDoesIt2012}. Following the success of our previous work on discriminating live speech from synthetic speech using learnable \gls{strf}s \cite{vuong2020learnable}, we  trained the \emph{STRFNet} system on \gls{sid} using the Librispeech \cite{panayotovLibrispeechASRCorpus2015} dataset
with artificially-added noise from Sound Bible\footnote{https://www.openslr.org/17/} to  learn the parameters of each Gabor-based \gls{strf} \cite{meyerRobustnessSpectrotemporalFeatures2011} automatically.  The \gls{sid} system was able to achieve an average of 95\% accuracy with 2484 speakers and signal-to-noise ratios (SNRs) ranging from 0 to 30 dB.  We then keep the learned \gls{strf}s fixed and utilize them for our loss.  We use 60 STRF kernels,  each with a time support of 300 milliseconds and a span of 20 channels on the Mel scale. This pipeline is depicted in the upper panel of \fref{fig:stme}.

\textbf{Loss function.}
Given $N$ \gls{strf}s, we define the response of the speech \gls{stftm} to the $i^{th}$ \gls{strf} to be
\begin{equation}
    \text{STMR}^{(i)}_S[t,k] = \text{STRF}^{(i)}[t,k] \star \log(m(|S[t,k]|^2)),
\end{equation}
where $m$ is a frequency integration function using the Mel weighting and $\star$ denotes cross-correlation. The loss function is then
\begin{equation}
    %\text{STME} = \frac{1}{N} \sum_{i=1}^N \frac{||\overrightarrow{\text{STMR}}^{(i)}_S - %\overrightarrow{\text{STMR}}^{(i)}_{\widehat{S}} ||^2}{||\overrightarrow{\text{STMR}}^{(i)}_S||^2},
    \text{STME} =  \frac{\sum_{i=1}^N ||\overrightarrow{\text{STMR}}^{(i)}_S - \overrightarrow{\text{STMR}}^{(i)}_{\widehat{S}} ||^2}{\sum_{i=1}^N ||\overrightarrow{\text{STMR}}^{(i)}_S||^2},
\end{equation}
where no time integration is applied to obtain the vector form of \gls{stmr}.
%The STME is similar to the STMI between the enhanced speech and clean speech as shown below.  
In general, the \gls{stme} is strongly motivated by \gls{stmi} and is in fact very similar to the weighted distance term in the definition of $\text{STMI}^{\text{T}}$ with a few differences. First, the \gls{stme} uses learned \gls{strf} kernels that are discriminatively trained to optimize for a \gls{sid} task. Second, the \gls{stmr} is calculated using cross-correlation instead of convolution. Third, the original auditory spectrogram \cite{chiMultiresolutionSpectrotemporalAnalysis2005} is approximated by the Mel-spectrogram and logarithmic compression. Finally, the \gls{stme} calculates a weighted \gls{mse} using instantaneous response. 

To train our enhancement system, we use the \gls{stme} loss in addition to the standard \gls{tfe} in \eref{eqn:mse_loss}. This is motivated by the observation that the \gls{stmr} is smooth and omits spectral details that are available in the \gls{tf} domain. We believe that the medium-time \gls{stme} loss complements the short-time \gls{tfe} for supervised \gls{se}.

A block diagram of the STME calculation is shown in the bottom panel of \fref{fig:stme}.  In the next section, we will describe the experimental setup used to evaluate the benefits of including the STME in the training loss.  We also assess the value of using automatically-learned Gabor-based STRF kernels compared to randomly-selected kernels.   

\section{Experimental Setup}

%As mentioned in the previous section, we learned the parameters to the Gabor-based STRF kernels by training  \emph{STRFNet} on speaker identification.  The system was trained using Librispeech\cite{panayotovLibrispeechASRCorpus2015} with noise randomly added from Sound Bible\cite{snyderMUSANMusicSpeech2015} and we followed the same training and test split used in \cite{ravanelliSpeakerRecognitionRaw2018}. 
% voicebank - Noisy speech database for training speech enhancement algorithms and TTS models
%demand - The Diverse Environments Multi-Channel Acoustic Noise Database (DEMAND): A database of multichannel environmental noise recordings
\textbf{Datasets.}
We used a small-scale and a large-scale dataset for evaluating the \gls{se} system. The small-scale dataset by Valentini \etal~\cite{valentini2016investigating} (VBD henceforth) contains 9.4 hours and 35 minutes of noisy speech in the training and test set, respectively. We downsampled the entire dataset to 16 kHz.  For the large-scale dataset, we used the Interspeech 2020 Deep Noise Suppression (DNS) dataset \cite{reddyINTERSPEECH2020Deep2020} with RIR responses provided by \cite{reddy2020icassp}.  
%The organizers provided clean speech from Librivox\footnote{https://librivox.org/} and noise clips from Audioset \cite{gemmekeAudioSetOntology2017}, Free Sound \cite{snyderMUSANMusicSpeech2015}, and DEMAND \cite{thiemann2013diverse}.  Additionally, we included simulated and real room impulse responses(RIR) from SLR26\footnote{http://www.openslr.org/26/} and SLR28\footnote{http://www.openslr.org/28/}, respectively. 
The DNS training set contains a total of 500 hours of noisy speech.  
%Noisy speech segments were generated by first convolving clean speech with a randomly sampled RIR and then adding randomly sampled noise at global SNRs of -5, 0, 5, 10, 15, 20, 25, 30 dB.  
For evaluation, the DNS dataset has two test sets named $no\_reverb$ and $with\_reverb$, which both contain 25 minutes of noisy speech.  In both datasets, we are also provided with the original clean speech that was used to artificially generate the noisy speech.  

We evaluated our system's capabilities to improve speaker verification performance in noisy conditions by using a modified version of the VoxCeleb1 test set \cite{nagraniVoxCelebLargeScaleSpeaker2017}.  The original test set contained 4874 speech pairs spoken by 40 unseen speakers.  We modified the VoxCeleb1 test set by randomly adding noise from the DNS test set at SNRs ranging from -6 to 6 dB.

%\subsection{Evaluation Metrics}
% pesq Perceptual evaluation of speech quality (PESQ)-a new method for speech quality assessment of telephone networks and codecs
% can add stoi
%We compared our proposed method with three different baselines by evaluating the objective speech quality of the enhanced speech and the equal error rate (EER) on speaker verification.  The objective speech quality metrics include the perceptual evaluation of speech quality (PESQ) \cite{rixPerceptualEvaluationSpeech2001}, scale-invariant signal-to-distortion ratio \cite{lerouxSDRHalfbakedWell2019}, and short-time objective intelligibility (STOI) \cite{taalShorttimeObjectiveIntelligibility2010} and were all used to evaluate the enhanced speech on the VBD and DNS dataset.  EER was used to evaluate how well the enhancement system improved a speaker verification system in noisy conditions.

\textbf{Training and evaluation procedure.}
To train our \gls{se} systems on the VBD dataset and DNS dataset, we randomly sampled 1-second noisy speech segments from the training data and 5-second noisy speech segments from the training data, respectively. All the \gls{se} systems were trained using the Adam optimizer with a learning rate of $5e^{-4}$ and a batch size of 64 in PyTorch. For evaluation, we used the perceptual evaluation of speech quality (PESQ) \cite{rixPerceptualEvaluationSpeech2001}, scale-invariant signal-to-distortion ratio \cite{lerouxSDRHalfbakedWell2019}, and short-time objective intelligibility (STOI) \cite{taalShorttimeObjectiveIntelligibility2010} metrics.  

To evaluate our \gls{se} systems on speaker verification, we used a DNN-based speaker verification system \cite{chungDefenceMetricLearning2020} pretrained on VoxCeleb2 \cite{chungVoxCeleb2DeepSpeaker2018}, a dataset that contains over 1 million utterances and 6112 speakers. The verification system obtained an equal error rate (EER) of 2.2\% on the VoxCeleb1 test set, which is one of the lowest reported EER compared to any other method with a similar number of parameters \cite{chungDefenceMetricLearning2020}.  Although the system was not trained with artificial degradation, all the audio clips were extracted from YouTube which will naturally contain different acoustical conditions. %With that being said,
%This is a suitable speaker verification system to evaluate our enhancement systems with.  
To test if our \gls{se} system improves the performance of this strong speaker verification system in noisy conditions, we added noise from the DNS test set to the original VoxCeleb1 test set at SNRs ranging from -6 to 6 dB.  We evaluated the EER of the speaker verification system on clips enhanced by the SE system.

%Additionally, all experiments were tracked using Weights \& Biases \cite{wandb} and PyTorch-Lightning \cite{falcon2019pytorch}.  

\textbf{Baseline systems.}
We evaluated three different baseline systems to illustrate the benefits of the additional STME loss.  Each of the baseline systems have the exact same network architecture, but they were each trained with different loss functions.  Our first baseline system, GRU(TFE), was trained only with the TFE loss.  To evaluate the benefits of the STME loss by itself, we trained a second baseline system, GRU(STME), using only the STME loss.  Our third baseline system, GRU(TFE+$\text{STME}^{R}$), was trained with both loss terms, although the parameters of the Gabor-based STRF kernels that were used to calculate the STME were randomly selected.  Specifically, the temporal and spectral modulation frequencies are uniformly sampled over $[0, 50)$ Hz and $[0, 0.5)$ cycles per channel, respectively.  This comparison allows us to assess the benefits of using automatically-learned Gabor-based STRF kernels to train the system, GRU(TFE+STME).

\section{Experimental Results and Discussion}In this section, we present and discuss results of our \gls{stme} loss on speech enhancement and speaker verification.

\begin{table}
\centering
\caption{Table summarizing the objective speech quality evaluation on the VBD test set}
\begin{tabular}{l|l|l|l}
Methods                        & PESQ~ & SI-SDR & STOI  \\
                               & (MOS) & (dB)   & (\%)  \\ 
\hline
Noisy                          & 1.97       & 8.5        & 92.1       \\
ERNN \cite{takeuchi2020real}                      & 2.54       &  ---      &  ---     \\
GRU(TFE)                       & 2.68       & \textbf{17.0}       &  93.3     \\
GRU(STME)                      & 2.78      & 14.4       &  93.1     \\
GRU(TFE+STME$^{R}$)         & 2.76       & 16.9        & 93.2       \\
GRU(TFE+STME)                & \textbf{2.82}       & \textbf{17.0}        & \textbf{93.8}     
\end{tabular}
\label{table:Valentini-results}
\vspace{-.5 cm}

\end{table}

\begin{table}
\scalebox{.87}{
\label{table:DNS-result}
\centering
\caption{Table summarizing the objective speech quality evaluation on the DNS $no\_reverb$  ($with\_reverb$) test set}
%\hskip-.7cm

\begin{tabular}{l|l|l|l}
Method                 & PESQ        & SI-SDR        & STOI         \\
                       & (MOS)       & (dB)          & (\%)         \\ 
\hline
Noisy                  & 1.58 (1.82) & 9.1 (9.0)   & 91.5 (86.6)  \\
DNS Baseline \cite{xiaWeightedSpeechDistortion2020} & 1.83 (1.52) & 12.5 (9.2)  & 90.6 (82.1)  \\
GRU(TFE)               & 2.27 (2.36) & 14.9 (13.2) & 94.2 (89.4)  \\
GRU(STME)              & 2.59 (2.64) & 12.4 (12.0) & 94.2 (90.1)  \\
GRU(TFE+STME$^{R}$)          & 2.57 (2.63) & \textbf{15.9 (14.5)} & 95.2 (90.6)  \\
GRU(TFE+STME)          & \textbf{2.71 (2.75)} & \textbf{15.9 (14.5)} & \textbf{95.5 (91.2)} 
\end{tabular}
}
\end{table}

\textbf{VBD results.}
In Table \ref{table:Valentini-results}, we show the objective evaluation of each SE system trained with different losses using the VBD dataset.  On a small dataset, our GRU(TFE+STME) outperformed the GRU(TFE) baseline in all the objective metrics.  Interestingly, different initialization of the STRF kernels in GRU(TFE+STME$^{R}$) resulted in similar improvements over the baseline TFE loss, but roughly 20\% of the time the random parameter selection resulted in a much worse performance.  This highlights the benefits of using automatically-learned Gabor-based STRF kernels over randomly-selected Gabor-based STRF kernels.    

\textbf{DNS results.}
The objective evaluation of our SE systems trained with different losses using the DNS  $no\_reverb$ and $with\_reverb$ test set is summarized in Table 2.  Even with a large amount of training data, our  GRU(TFE+STME) loss function outperformed both the GRU(TFE) baseline and the provided challenge baseline \cite{xiaWeightedSpeechDistortion2020} in all the objective metrics.  Most notably, there is a significant improvement in PESQ which results in our system having a similar PESQ to the top system in the official DNS challenge \cite{isik2020poconet}.  We also evaluated the benefits of the \gls{stme} loss by itself, GRU(STME).  Curiously, training with only our the \gls{stme} loss provides a higher PESQ but much lower SI-SDR compared to training with only the TFE loss.  Nevertheless,  optimizing the combination of both losses  during training caused both the PESQ and SI-SDR scores to increase compared to training on each loss individually. This confirms our belief that the medium-time \gls{stme} loss is complemented by the short-time \gls{tfe} loss. As in the VBD experiments, the use of  automatically-learned Gabor-based STRF kernels provides a greater increase of PESQ scores compared to the use of randomly-selected Gabor-based STRF kernels.
\begin{figure}
\scalebox{.65}{
\centering
  \includegraphics[width=5 in]{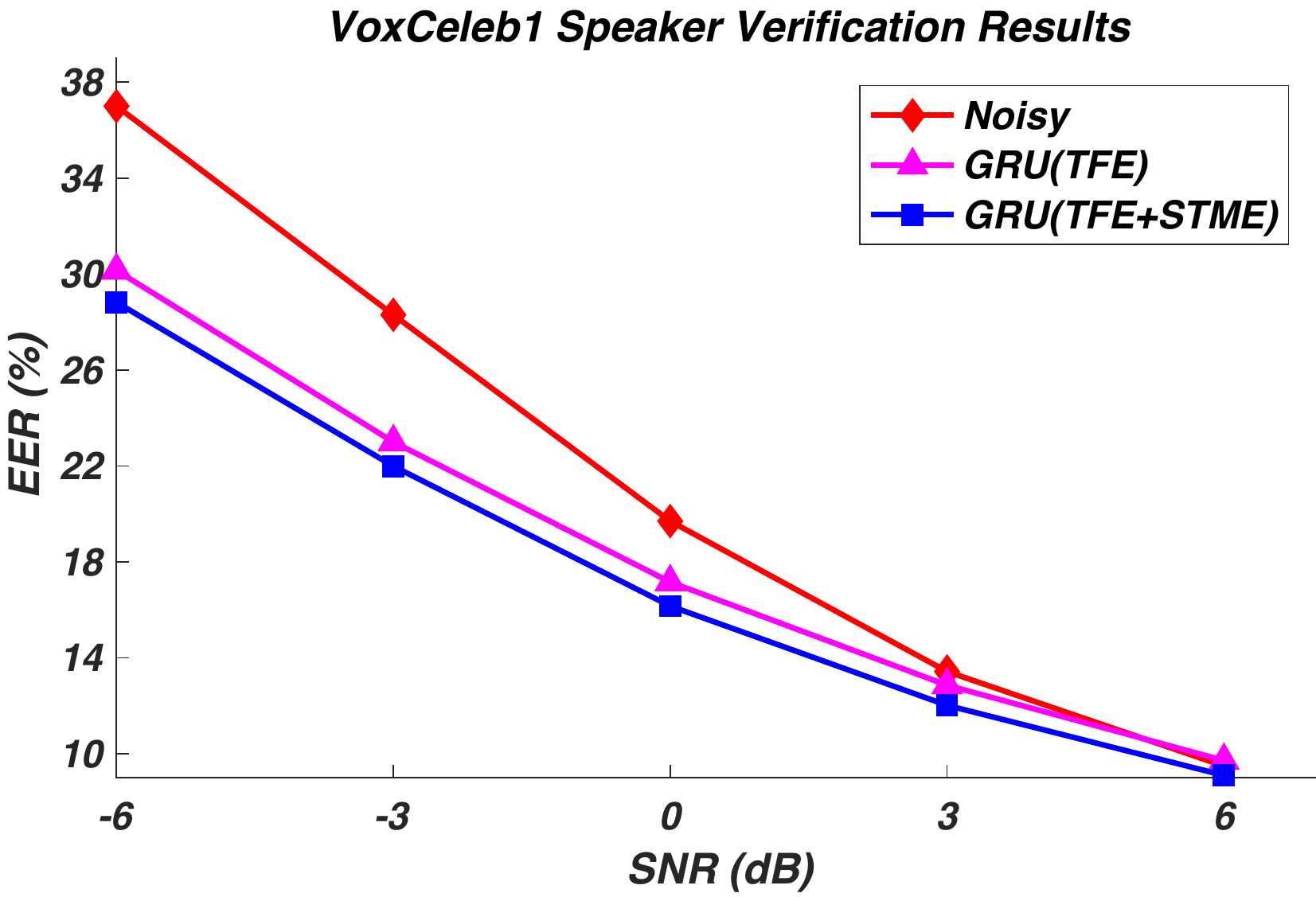}
  \caption{Equal error rates on the VoxCeleb1 Test Set.}
  %\caption{Experimental results.  Left panel: Error rates from ASVspoof-LA task.  Right panel: EER and DCF scores for VTD task. }
  \label{fig:EER_ASV}
  %\vspace{-3 mm}

  }
  \vspace{-5mm}
\end{figure}

\textbf{Speaker verification results.}
The performance of the speaker verification system with noise added at low SNRs is shown in Figure \ref{fig:EER_ASV}.  At low SNRs, the speaker verification system's performance starts to substantially degrade.  Our system GRU(TFE+STME) improved the EER by an average of 15.4\% relative and outperformed our baseline GRU(TFE) by an average of 5.5\% relative.  At higher SNRs, the verification system's performance quickly approached the state-of-the-art performance of 2.2\% and our enhancement systems did not provide any additional benefits. 
\section{Conclusions}
In this paper, we introduced a novel modulation-domain loss for training neural-network-based speech enhancement systems.  We showed that by adding   spectro-temporal modulation error to the standard time-frequency error during training, all three common objective speech quality metrics substantially improved on two different datasets.  Additionally, we demonstrated the value of utilizing automatically-learned Gabor-based STRF kernels over randomly-selected kernels.  We also showed that our speech enhancement system can improve a strong speaker verification system at low SNRs.  In the future, we plan on exploring deep-learning-based techniques to perform SE directly in the modulation domain.  We will also explore ways of directly optimizing the STRF parameters for speech enhancement.

\vfill\pagebreak

% References should be produced using the bibtex program from suitable
% BiBTeX files (here: strings, refs, manuals). The IEEEbib.bst bibliography
% style file from IEEE produces unsorted bibliography list.
% -------------------------------------------------------------------------
\bibliographystyle{IEEEtran}

\patchcmd{\thebibliography}
  {\settowidth}
  {\setlength{\itemsep}{0pt plus 0.1pt}\settowidth}
  {}{}
\apptocmd{\thebibliography}
  {\footnotesize}
  {}{}    
\bibliography{ICASSP2021}

% Generated by IEEEtran.bst, version: 1.13 (2008/09/30)
\begin{thebibliography}{10}
\providecommand{\url}[1]{#1}
\csname url@samestyle\endcsname
\providecommand{\newblock}{\relax}
\providecommand{\bibinfo}[2]{#2}
\providecommand{\BIBentrySTDinterwordspacing}{\spaceskip=0pt\relax}
\providecommand{\BIBentryALTinterwordstretchfactor}{4}
\providecommand{\BIBentryALTinterwordspacing}{\spaceskip=\fontdimen2\font plus
\BIBentryALTinterwordstretchfactor\fontdimen3\font minus
  \fontdimen4\font\relax}
\providecommand{\BIBforeignlanguage}[2]{{%
\expandafter\ifx\csname l@#1\endcsname\relax
\typeout{** WARNING: IEEEtran.bst: No hyphenation pattern has been}%
\typeout{** loaded for the language `#1'. Using the pattern for}%
\typeout{** the default language instead.}%
\else
\language=\csname l@#1\endcsname
\fi
#2}}
\providecommand{\BIBdecl}{\relax}
\BIBdecl

\bibitem{luoConvTasNetSurpassingIdeal2019}
Y.~Luo and N.~Mesgarani, ``\BIBforeignlanguage{en}{Conv-{{TasNet}}: Surpassing
  ideal time-frequency magnitude masking for speech separation},''
  \emph{\BIBforeignlanguage{en}{IEEE/ACM Transactions on Audio, Speech, and
  Language Processing}}, vol.~27, no.~8, pp. 1256--1266, Aug. 2019.

\bibitem{loizouReasonsWhyCurrent2011}
P.~C. Loizou and G.~Kim, ``Reasons why current speech-enhancement algorithms do
  not improve speech intelligibility and suggested solutions,'' \emph{IEEE/ACM
  Transactions on Audio, Speech, and Language Processing}, vol.~19, no.~1, pp.
  47--56, Jan. 2011.

\bibitem{kryterValidationArticulationIndex1962}
K.~D. Kryter, ``\BIBforeignlanguage{en}{Validation of the articulation
  index},'' \emph{\BIBforeignlanguage{en}{The Journal of the Acoustical Society
  of America}}, vol.~34, no.~11, pp. 1698--1702, Nov. 1962.

\bibitem{wangTrainingTargetsSupervised2014}
Y.~Wang, A.~Narayanan, and {DeLiang Wang}, ``On training targets for supervised
  speech separation,'' \emph{IEEE/ACM Transactions on Audio, Speech, and
  Language Processing}, vol.~22, no.~12, pp. 1849--1858, Dec. 2014.

\bibitem{braunConsolidatedViewLoss2020}
S.~Braun and I.~Tashev, ``\BIBforeignlanguage{en}{A consolidated view of loss
  functions for supervised deep learning-based speech enhancement},'' Sep.
  2020.

\bibitem{ephraimSpeechEnhancementUsing1984}
Y.~Ephraim and D.~Malah, ``Speech enhancement using a minimum-mean square error
  short-time spectral amplitude estimator,'' \emph{IEEE Transactions on
  Acoustics, Speech, and Signal Processing}, vol.~32, no.~6, pp. 1109--1121,
  Dec. 1984.

\bibitem{ephraimSpeechEnhancementUsing1985}
------, ``\BIBforeignlanguage{en}{Speech enhancement using a minimum
  mean-square error log-spectral amplitude estimator},''
  \emph{\BIBforeignlanguage{en}{IEEE Transactions on Acoustics, Speech, and
  Signal Processing}}, vol.~33, no.~2, pp. 443--445, Apr. 1985.

\bibitem{limEnhancementBandwidthCompression1979}
J.~S. Lim and A.~V. Oppenheim, ``Enhancement and bandwidth compression of noisy
  speech,'' \emph{Proceedings of the IEEE}, vol.~67, no.~12, pp. 1586--1604,
  Dec. 1979.

\bibitem{loizouSpeechEnhancementTheory2013}
P.~C. Loizou, \emph{\BIBforeignlanguage{en}{Speech Enhancement : Theory and
  Practice, Second Edition}}.\hskip 1em plus 0.5em minus 0.4em\relax {CRC
  Press}, Feb. 2013.

\bibitem{martin-donasDeepLearningLoss2018}
J.~M. {Martin-Donas}, A.~M. Gomez, J.~A. Gonzalez, and A.~M. Peinado,
  ``\BIBforeignlanguage{en}{A deep learning loss function based on the
  perceptual evaluation of the speech quality},''
  \emph{\BIBforeignlanguage{en}{IEEE Signal Processing Letters}}, vol.~25,
  no.~11, pp. 1680--1684, Nov. 2018.

\bibitem{zhaoPerceptuallyGuidedSpeech2018}
Y.~Zhao, B.~Xu, R.~Giri, and T.~Zhang, ``\BIBforeignlanguage{en}{Perceptually
  guided speech enhancement using deep neural networks},'' in
  \emph{\BIBforeignlanguage{en}{2018 {{IEEE International Conference}} on
  {{Acoustics}}, {{Speech}} and {{Signal Processing}} ({{ICASSP}})}}.\hskip 1em
  plus 0.5em minus 0.4em\relax {Calgary, AB}: {IEEE}, Apr. 2018, pp.
  5074--5078.

\bibitem{steeneken1980physical}
H.~J. Steeneken and T.~Houtgast, ``A physical method for measuring
  speech-transmission quality,'' \emph{The Journal of the Acoustical Society of
  America}, vol.~67, no.~1, pp. 318--326, 1980.

\bibitem{paytonMethodDetermineSpeech1999}
K.~L. Payton and L.~D. Braida, ``A method to determine the speech transmission
  index from speech waveforms,'' \emph{The Journal of the Acoustical Society of
  America}, vol. 106, no.~6, pp. 3637--3648, Nov. 1999.

\bibitem{elhilaliSpectrotemporalModulationIndex2003}
M.~Elhilali, T.~Chi, and S.~A. Shamma, ``\BIBforeignlanguage{en}{A
  spectro-temporal modulation index ({{STMI}}) for assessment of speech
  intelligibility},'' \emph{\BIBforeignlanguage{en}{Speech Communication}},
  vol.~41, no. 2-3, pp. 331--348, Oct. 2003.

\bibitem{mesgaraniDenoisingDomainSpectrotemporal2007}
N.~Mesgarani and S.~Shamma, ``\BIBforeignlanguage{en}{Denoising in the domain
  of spectrotemporal modulations},'' \emph{\BIBforeignlanguage{en}{EURASIP
  Journal on Audio, Speech, and Music Processing}}, vol. 2007, pp. 1--8, 2007.

\bibitem{mirbagheriNonlinearFilteringSpectrotemporal2010}
M.~Mirbagheri, N.~Mesgarani, and S.~Shamma, ``Nonlinear filtering of
  spectrotemporal modulations in speech enhancement,'' in \emph{2010 {{IEEE
  International Conference}} on {{Acoustics}}, {{Speech}} and {{Signal
  Processing}} (ICASSP)}.\hskip 1em plus 0.5em minus 0.4em\relax {Dallas, TX,
  USA}: {IEEE}, Mar. 2010, pp. 5478--5481.

\bibitem{vuong2020learnable}
T.~Vuong, Y.~Xia, and R.~M. Stern, ``Learnable spectro-temporal receptive
  fields for robust voice type discrimination,'' in \emph{Interspeech
  2020}.\hskip 1em plus 0.5em minus 0.4em\relax {ISCA}, Oct. 2020, pp.
  1957--1961.

\bibitem{chiMultiresolutionSpectrotemporalAnalysis2005}
T.~Chi, P.~Ru, and S.~A. Shamma, ``\BIBforeignlanguage{en}{Multiresolution
  spectrotemporal analysis of complex sounds},''
  \emph{\BIBforeignlanguage{en}{The Journal of the Acoustical Society of
  America}}, vol. 118, no.~2, pp. 887--906, Aug. 2005.

\bibitem{mesgaraniDiscriminationSpeechNonspeech2006}
N.~Mesgarani, M.~Slaney, and S.~Shamma,
  ``\BIBforeignlanguage{en}{Discrimination of speech from nonspeech based on
  multiscale spectro-temporal modulations},''
  \emph{\BIBforeignlanguage{en}{IEEE/ACM Transactions on Audio, Speech and
  Language Processing}}, vol.~14, no.~3, pp. 920--930, May 2006.

\bibitem{meyerRobustnessSpectrotemporalFeatures2011}
B.~T. Meyer and B.~Kollmeier, ``\BIBforeignlanguage{en}{Robustness of
  spectro-temporal features against intrinsic and extrinsic variations in
  automatic speech recognition},'' \emph{\BIBforeignlanguage{en}{Speech
  Communication}}, vol.~53, no.~5, pp. 753--767, May 2011.

\bibitem{ravuriEasyDoesIt2012}
S.~V. Ravuri and N.~Morgan, ``\BIBforeignlanguage{en}{Easy does it: Robust
  spectro-temporal many-stream {{ASR}} without fine tuning streams},'' in
  \emph{\BIBforeignlanguage{en}{2012 {{IEEE International Conference}} on
  {{Acoustics}}, {{Speech}} and {{Signal Processing}} ({{ICASSP}})}}.\hskip 1em
  plus 0.5em minus 0.4em\relax {Kyoto, Japan}: {IEEE}, Mar. 2012, pp.
  4309--4312.

\bibitem{xiaWeightedSpeechDistortion2020}
Y.~Xia, S.~Braun, C.~K.~A. Reddy, H.~Dubey, R.~Cutler, and I.~Tashev,
  ``Weighted speech distortion losses for neural-network-based real-time speech
  enhancement,'' in \emph{2020 {{IEEE International Conference}} on
  {{Acoustics}}, {{Speech}} and {{Signal Processing}} ({{ICASSP}})}.\hskip 1em
  plus 0.5em minus 0.4em\relax {IEEE}, May 2020, pp. 871--875.

\bibitem{panayotovLibrispeechASRCorpus2015}
V.~Panayotov, G.~Chen, D.~Povey, and S.~Khudanpur, ``Librispeech: {{An ASR}}
  corpus based on public domain audio books,'' in \emph{2015 {{IEEE
  International Conference}} on {{Acoustics}}, {{Speech}} and {{Signal
  Processing}} ({{ICASSP}})}.\hskip 1em plus 0.5em minus 0.4em\relax {IEEE},
  Apr. 2015, pp. 5206--5210.

\bibitem{valentini2016investigating}
C.~Valentini-Botinhao, X.~Wang, S.~Takaki, and J.~Yamagishi, ``Investigating
  {{RNN}}-based speech enhancement methods for noise-robust text-to-speech.''
  in \emph{SSW}, 2016, pp. 146--152.

\bibitem{reddyINTERSPEECH2020Deep2020}
C.~K.~A. Reddy, E.~Beyrami, H.~Dubey, V.~Gopal, R.~Cheng, R.~Cutler,
  S.~Matusevych, R.~Aichner, A.~Aazami, S.~Braun, P.~Rana, S.~Srinivasan, and
  J.~Gehrke, ``The {{INTERSPEECH}} 2020 deep noise suppression challenge:
  Datasets, subjective speech quality and testing framework,''
  \emph{arXiv:2001.08662 [cs, eess]}, Apr. 2020.

\bibitem{reddy2020icassp}
C.~K.~A. Reddy, H.~Dubey, V.~Gopal, R.~Cutler, S.~Braun, H.~Gamper, R.~Aichner,
  and S.~Srinivasan, ``{{ICASSP}} 2021 deep noise suppression challenge,''
  2020.

\bibitem{nagraniVoxCelebLargeScaleSpeaker2017}
A.~Nagrani, J.~S. Chung, and A.~Zisserman,
  ``\BIBforeignlanguage{en}{{{VoxCeleb}}: A large-scale speaker identification
  dataset},'' in \emph{\BIBforeignlanguage{en}{Interspeech 2017}}.\hskip 1em
  plus 0.5em minus 0.4em\relax {ISCA}, Aug. 2017, pp. 2616--2620.

\bibitem{rixPerceptualEvaluationSpeech2001}
A.~Rix, J.~Beerends, M.~Hollier, and A.~Hekstra, ``Perceptual evaluation of
  speech quality ({{PESQ}})-a new method for speech quality assessment of
  telephone networks and codecs,'' in \emph{2001 {{IEEE International
  Conference}} on {{Acoustics}}, {{Speech}}, and {{Signal Processing}}
  ({{ICASSP}})}, vol.~2.\hskip 1em plus 0.5em minus 0.4em\relax {Salt Lake
  City, UT, USA}: {IEEE}, 2001, pp. 749--752.

\bibitem{lerouxSDRHalfbakedWell2019}
J.~Le~Roux, S.~Wisdom, H.~Erdogan, and J.~R. Hershey, ``{{SDR}}\textendash
  half-baked or well done?'' in \emph{2019 {{IEEE International Conference}} on
  {{Acoustics}}, {{Speech}} and {{Signal Processing}} ({{ICASSP}})}.\hskip 1em
  plus 0.5em minus 0.4em\relax {IEEE}, 2019, pp. 626--630.

\bibitem{taalShorttimeObjectiveIntelligibility2010}
C.~H. Taal, R.~C. Hendriks, R.~Heusdens, and J.~Jensen,
  ``\BIBforeignlanguage{en}{A short-time objective intelligibility measure for
  time-frequency weighted noisy speech},'' in
  \emph{\BIBforeignlanguage{en}{2010 {{IEEE International Conference}} on
  {{Acoustics}}, {{Speech}} and {{Signal Processing}} ({{ICASSP}})}}.\hskip 1em
  plus 0.5em minus 0.4em\relax {Dallas, TX, USA}: {IEEE}, 2010, pp. 4214--4217.

\bibitem{chungDefenceMetricLearning2020}
J.~S. Chung, J.~Huh, S.~Mun, M.~Lee, H.~S. Heo, S.~Choe, C.~Ham, S.~Jung, B.-J.
  Lee, and I.~Han, ``In defence of metric learning for speaker recognition,''
  \emph{arXiv:2003.11982 [cs, eess]}, Apr. 2020.

\bibitem{chungVoxCeleb2DeepSpeaker2018}
J.~S. Chung, A.~Nagrani, and A.~Zisserman,
  ``\BIBforeignlanguage{en}{{{VoxCeleb2}}: Deep speaker recognition},'' in
  \emph{\BIBforeignlanguage{en}{Interspeech 2018}}.\hskip 1em plus 0.5em minus
  0.4em\relax {ISCA}, Sep. 2018, pp. 1086--1090.

\bibitem{takeuchi2020real}
D.~Takeuchi, K.~Yatabe, Y.~Koizumi, Y.~Oikawa, and N.~Harada, ``Real-time
  speech enhancement using equilibriated {{RNN}},'' in \emph{2020 {{IEEE
  International Conference}} on {{Acoustics}}, {{Speech}} and {{Signal
  Processing}} ({{ICASSP}})}.\hskip 1em plus 0.5em minus 0.4em\relax IEEE,
  2020, pp. 851--855.

\bibitem{isik2020poconet}
U.~Isik, R.~Giri, N.~Phansalkar, J.-M. Valin, K.~Helwani, and A.~Krishnaswamy,
  ``Poconet: Better speech enhancement with frequency-positional embeddings,
  semi-supervised conversational data, and biased loss,'' \emph{arXiv preprint
  arXiv:2008.04470}, 2020.

\end{thebibliography}

\end{document}